\documentclass[journal,twoside,web]{ieeecolor}
\usepackage{tmi}
\usepackage{cite}
\usepackage{amsmath,amssymb,amsfonts}
\usepackage{algorithmic}
\usepackage{graphicx}
\usepackage{textcomp}

\usepackage{comment}
\usepackage{multirow}

\def\BibTeX{{\rm B\kern-.05em{\sc i\kern-.025em b}\kern-.08em
    T\kern-.1667em\lower.7ex\hbox{E}\kern-.125emX}}
\begin{document}
\bstctlcite{IEEEexample:BSTcontrol}

\title{Modeling 3D cardiac contraction and relaxation with point cloud deformation networks}
\author{Marcel Beetz, Abhirup Banerjee, \IEEEmembership{Member, IEEE}, and Vicente Grau
\thanks{Manuscript received -- July 2023. This work was supported in part by the Stiftung der Deutschen Wirtschaft (Foundation of German Business), the Royal Society (Grant No. URF{\textbackslash}R1{\textbackslash}221314), the British Heart Foundation (Grant No. PG/20/21/35082), and the CompBioMed 2 Centre of Excellence in Computational Biomedicine (European Commission Horizon 2020 research and innovation programme, grant agreement No. 823712). (Corresponding author: Marcel Beetz.)}
\thanks{M. Beetz and V. Grau are with the Institute of Biomedical Engineering, Department of Engineering Science, University of Oxford, Oxford OX3 7DQ, UK (e-mail: \{marcel.beetz,vicente.grau\}@eng.ox.ac.uk).}
\thanks{A. Banerjee is with the Institute of Biomedical Engineering, Department of Engineering Science, University of Oxford, Oxford OX3 7DQ, UK, and also with the Division of Cardiovascular Medicine, Radcliffe Department of Medicine, University of Oxford, Oxford OX3 9DU, UK (e-mail: abhirup.banerjee@eng.ox.ac.uk).}}

\maketitle

\begin{abstract}
Global single-valued biomarkers of cardiac function  typically used in clinical practice, such as ejection fraction, provide limited insight on the true 3D cardiac deformation process and hence, limit the understanding of both healthy and pathological cardiac mechanics. In this work, we propose the Point Cloud Deformation Network (PCD-Net) as a novel geometric deep learning approach to model 3D cardiac contraction and relaxation between the extreme ends of the cardiac cycle. It employs the recent advances in point cloud-based deep learning into an encoder-decoder structure, in order to enable efficient multi-scale feature learning directly on multi-class 3D point cloud representations of the cardiac anatomy. We evaluate our approach on a large dataset of over 10,000 cases from the UK Biobank study and find average Chamfer distances between the predicted and ground truth anatomies below the pixel resolution of the underlying image acquisition. Furthermore, we observe similar clinical metrics between predicted and ground truth populations and show that the PCD-Net can successfully capture subpopulation-specific differences between normal subjects and myocardial infarction (MI) patients. We then demonstrate that the learned 3D deformation patterns outperform multiple clinical benchmarks by 13\% and 7\% in terms of area under the receiver operating characteristic curve for the tasks of prevalent MI detection and incident MI prediction and by 7\% in terms of Harrell’s concordance index for MI survival analysis.
\end{abstract}


\section{Introduction}
\label{sec:introduction}
The human heart acts as a mechanical pump, which contracts and expands in a cyclical pattern to ensure adequate blood supply throughout the body via the circulatory system. In clinical practice, the mechanical function of the heart is typically assessed with the help of image-based biomarkers, such as ejection fraction. While such metrics enable a single-value global evaluation, they cannot capture the complex non-linear 3D deformation patterns of the underlying cardiac mechanics. This places limitations on the understanding and accurate diagnosis of a variety of diseases, including myocardial infarction (MI), which have previously been shown to benefit  from the use of more intricate and localized shape biomarkers \cite{acero2022understanding,suinesiaputra2017statistical}.

Considerable research efforts have focused on creating more accurate 3D models of cardiac mechanics. Early approaches in this direction typically employed classical numerical optimization techniques, such as finite element methods, on 3D mesh representations of the cardiac anatomy to solve systems of differential equations that govern 3D cardiac deformation \cite{sugiura2012multi,krishnamurthy2013patient,lopez2015three,chabiniok2016multiphysics,hong2019modeling}. Hereby, the input 3D meshes either consist of virtual models that approximate cardiac anatomy in a general sense or are obtained directly from medical images to enable personalized simulations \cite{marchesseau2015personalization,duchateau2017model}. As the mesh generation step often comes with a significant computational cost and possible errors, meshless simulation approaches frequently relying on point cloud representations have also been proposed \cite{wong2010meshfree,naceur2015efficient,lluch2019breaking,lluch2020calibration}. 

More recently, a variety of machine learning and deep learning techniques have been used to model 3D cardiac anatomy and deformation with cardiac morphology represented as either 2D images, 3D voxel grids, graphs, meshes, or point clouds \cite{duchateau2020machine,arzani2022machine}. Corral Acero et~al. \cite{acero2022understanding} employed principal component analysis to capture the deformation of 3D surface meshes between the end-diastolic (ED) and end-systolic (ES) phases of the cardiac cycle, while Di Folco et~al. \cite{pamela2022characterizing} developed a manifold learning approach to find joint embeddings of myocardial shape and deformation. Several deep learning approaches for 2D or 3D image data have also been explored, including probabilistic motion models to simulate deformation patterns for cardiac image sequences \cite{krebs2019probabilistic}, vertex-wise trajectories of downsampled cardiac meshes across the cardiac cycle to train a time series autoencoder for survival prediction \cite{bello2019deep}, combining a Siamese network style motion estimation branch with a segmentation branch for joint segmentation and cardiac motion estimation in cardiac MR images \cite{qin2018joint}, predicting volume and strain parameters of cardiac mechanics from cine MR images \cite{morales2021deepstrain}, generative adversarial networks to predict cardiac contraction in 2D images between the extreme ends of the cardiac cycle \cite{Ossenberg-Engels2019}, etc. Graph neural networks have also been used to model left ventricular (LV) motion on 2D cardiac image slices \cite{lu2020modelling,lu2021multiscale} and cardiac meshes \cite{meng2022mesh,beetz2023mesh,beetz2022interpretable,beetz2023post}, or emulate numerical solvers of cardiac mechanics on a virtual 3D mesh of the LV \cite{dalton2021graph}. Point clouds are another common data format used to represent 3D surfaces and encode the information as a set of point coordinates. This straightforward representation makes point clouds particularly easy to work with and considerably more efficient at storing and processing surface-level data, especially compared to voxelgrids whose $O(N^{3})$ space complexity limits data resolution and representation accuracy. Points clouds also offer high flexibility, particularly in comparison to meshes, as they do not require any explicit encoding of connectivity information between points nor any shared vertex correspondence across the dataset, thus reducing the need for complex local registration procedures or potentially error-prone meshing steps. Despite these advantages, point cloud deep learning has only seen limited usage for 3D cardiac modeling tasks, such as 3D shape reconstruction \cite{zhou2019one,beetz2021biventricular,beetz2023point2mesh}, pathology classification \cite{chang2020automatic,beetz20233d}, mechanics modeling \cite{beetz2021predicting}, 3D anatomy modeling \cite{beetz2021generating}, and the combined modeling of cardiac anatomy and electrophysiology \cite{beetz2022multi,beetz2022combined,li2022deep}.

In this work, we propose the Point Cloud Deformation Network (PCD-Net) as the first point cloud-based deep learning approach to predict 3D cardiac deformation between the two ends of the cardiac cycle. Its architecture follows a hierarchical encoder-decoder design based on recent advances in geometric deep learning, enabling efficient multi-scale feature learning on a high-resolution point cloud representation of the biventricular anatomy. Compared to previous approaches, the PCD-Net is not limited to 2D slice representations \cite{Ossenberg-Engels2019,lu2020modelling,lu2021multiscale}, is tasked to predict 3D cardiac shapes between different phases of the cardiac cycle as opposed to simply encode motion inputs \cite{bello2019deep,acero2022understanding,pamela2022characterizing}, uses point clouds instead of voxelgrids or meshes to represent the cardiac anatomy \cite{Ossenberg-Engels2019,lu2021multiscale,acero2022understanding,meng2022mesh}, and is trained and evaluated on a large dataset of 10,000 real cases as opposed to smaller numbers of real \cite{bello2019deep,krebs2019probabilistic,morales2021deepstrain} or synthetic subjects \cite{dalton2021graph}. Furthermore, the PCD-Net can incorporate information from multiple cardiac substructures \cite{dalton2021graph,morales2021deepstrain,meng2022mesh}, has been applied to pathological cases in addition to healthy subjects \cite{dalton2021graph,lu2021multiscale,meng2022mesh}, is validated for both cardiac contraction and relaxation \cite{Ossenberg-Engels2019}, uses high-resolution representations of the cardiac surface \cite{bello2019deep,lu2020modelling}, and contains an interpretable low-dimensional representation of cardiac deformation.

In summary, our key contributions are as follows:
\begin{itemize}
    \item We present a novel geometric deep learning approach, the PCD-Net, for direct modeling of 3D cardiac mechanics of the biventricular anatomy between the extreme ends of the cardiac cycle;
    \item We successfully integrate it into a multi-step pipeline to enable a fast and fully automatic application to raw cine MR images;
    \item We assess the PCD-Net's predictive ability for both 3D cardiac contraction and relaxation on a large cine MRI dataset of over 10,000 real subjects;
    \item We evaluate the predicted 3D shapes from a clinical perspective using widely used image-based biomarkers for cardiac anatomy and function;
    \item We show the proposed method's capability to capture subpopulation-specific differences between control and pathological cases; and
    \item We demonstrate the utility of the low-dimensional latent space representation of healthy cardiac deformation learned by the network for the detection of prevalent and incident MI as well as survival analysis.
\end{itemize}

\section{Methods}

\subsection{Overview}

\begin{figure*}[htbp]
	\begin{minipage}[b]{1.0\linewidth}
		\centering
		\centerline{\includegraphics[width=1.0\textwidth]{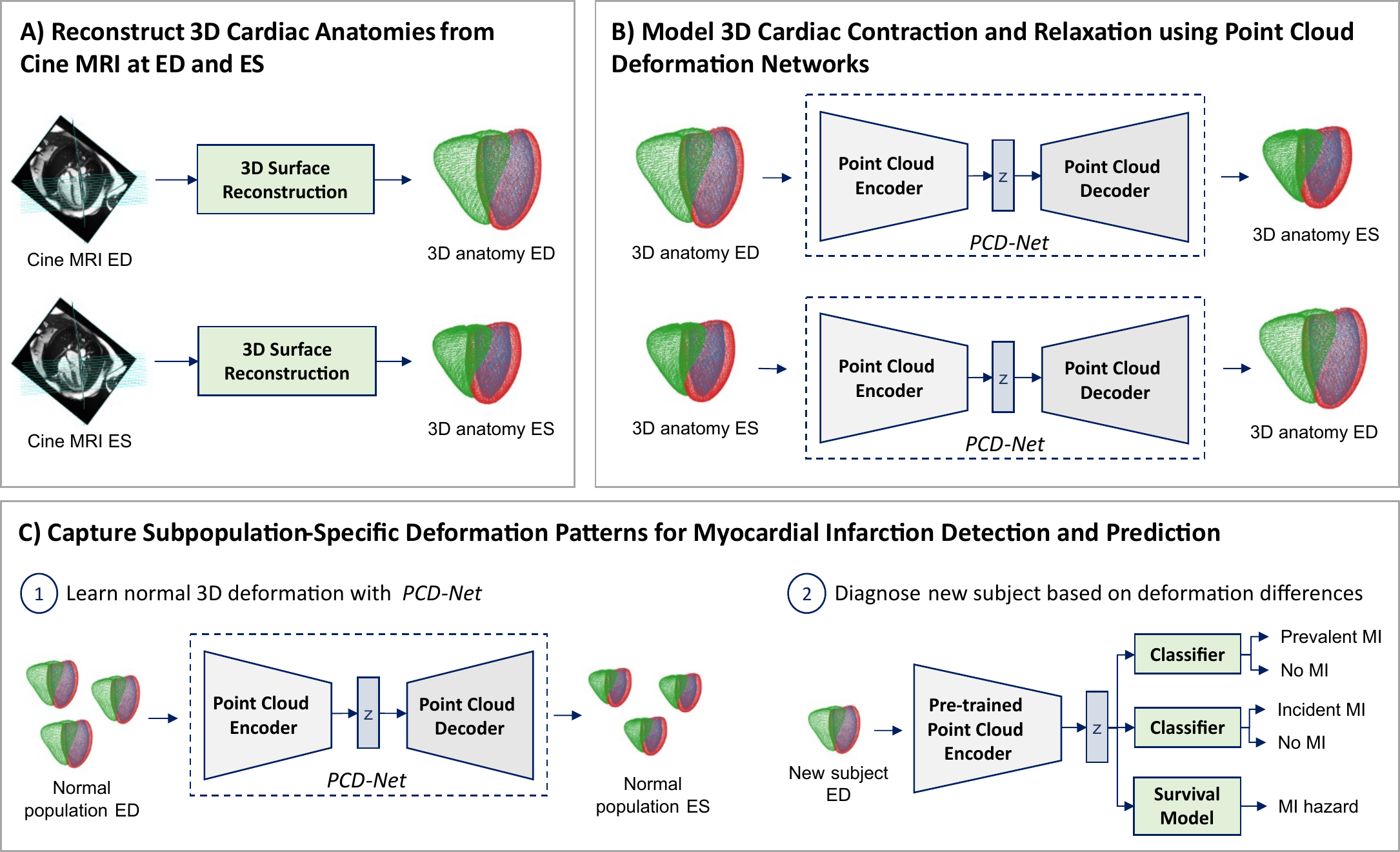}}
	\end{minipage}
	\caption{Overview of the proposed pipeline for cardiac deformation modeling. It first reconstructs surface models of the 3D cardiac anatomy in the form of multi-class point clouds from cine MR images at both the ED and ES phases using a fully automated process (A). The reconstructed shapes are then used as inputs for two separate PCD-Nets, one to predict cardiac contraction (ED to ES) and the other to predict cardiac relaxation (ES to ED). By training on a large population of healthy cases, the PCD-Nets learn typical phenotypes of ``normal" 3D deformation for given input shapes, which can then be used to identify pathological deviations in cardiac mechanics of new subjects (C). }
	\label{fig:overview}
\end{figure*}

This work presents the PCD-Net as a novel approach to model 3D mechanical deformation of the heart and embeds it into a multi-step pipeline to enable a direct application to cine MR images (Fig.~\ref{fig:overview}).

To this end, we first obtain 3D representations of the biventricular anatomy at the ED and ES phases of the cardiac cycle in the form of high-resolution multi-class point clouds (Fig.~\ref{fig:overview}-A) (Sec.~\ref{ssec:dataset}). We then train two separate PCD-Nets, the first to predict the ES anatomy based on the ED anatomy, and the second to predict ED from ES. This allows us to model both cardiac contraction and relaxation and investigate overall performance on a large number of both normal and pathological cases (Fig.~\ref{fig:overview}-B). Next, we retrain the two PCD-Nets on a large population of exclusively healthy subjects to capture accurate representations of ``normal" 3D cardiac contraction and relaxation (Fig.~\ref{fig:overview}-C-1). These learned normal deformation patterns are then shown to exhibit subpopulation-specific differences between healthy and MI cases (Sec.~\ref{ssec:exp_subpopulation_diffs}) and to be useful for outcome classification (Sec.~\ref{ssec:exp_mi_detection}) and survival analysis (Sec.~\ref{ssec:exp_mi_survival_pred}) (Fig.~\ref{fig:overview}-C-2).

\subsection{Dataset and Preprocessing}
\label{ssec:dataset}
The dataset used in this work is based on cine MR images of 10,237 subjects from the UK Biobank study \cite{petersen2015uk}. The same balanced steady-state free precession (bSSFP) protocol was consistently applied for all acquisitions with a voxel resolution of $1.8 \times 1.8 \times 8.0~\mbox{mm}^3$ for short-axis images and $1.8 \times 1.8 \times 6.0~\mbox{mm}^3$ for long-axis images \cite{petersen2013imaging}.

At the time of imaging, 294 subjects had survived a previous MI (prevalent MI). 235 subjects had an MI diagnosis after imaging (incident MI), and no MI event was recorded for the remaining cases. All subjects with MI events and the corresponding MI dates were identified based on UK Biobank field ID 42000 (Date of myocardial infarction). Out of the remaining subjects, 5373 did not suffer from any cardiovascular disease or other common pathologies of the UK Biobank study as identified by UK Biobank field ID 20002 (see Table~\ref{tab:uk_biobank_cardiac_disease_codes}). We consider these cases as the control group in this study.

For each case, we first select the two-chamber long-axis (LAX) slice, the four-chamber LAX slice, and the short-axis stack at both the ED and ES phases of the cardiac cycle \cite{banerjee2021ptrsa,banerjee2021miua}. We segment these and use them to reconstruct 3D surface models of the biventricular anatomies in the form of multi-class point clouds using the multi-step pipeline described in \cite{beetz2023multi,beetz2021biventricular}. We encode three cardiac substructures, namely the LV endocardium, LV epicardium, and right ventricular (RV) endocardium, as separate classes in the reconstructed point clouds.

\subsection{Point Cloud Deformation Network Architecture}

\begin{figure}[htbp]
	\centerline{\includegraphics[width=0.495\textwidth]{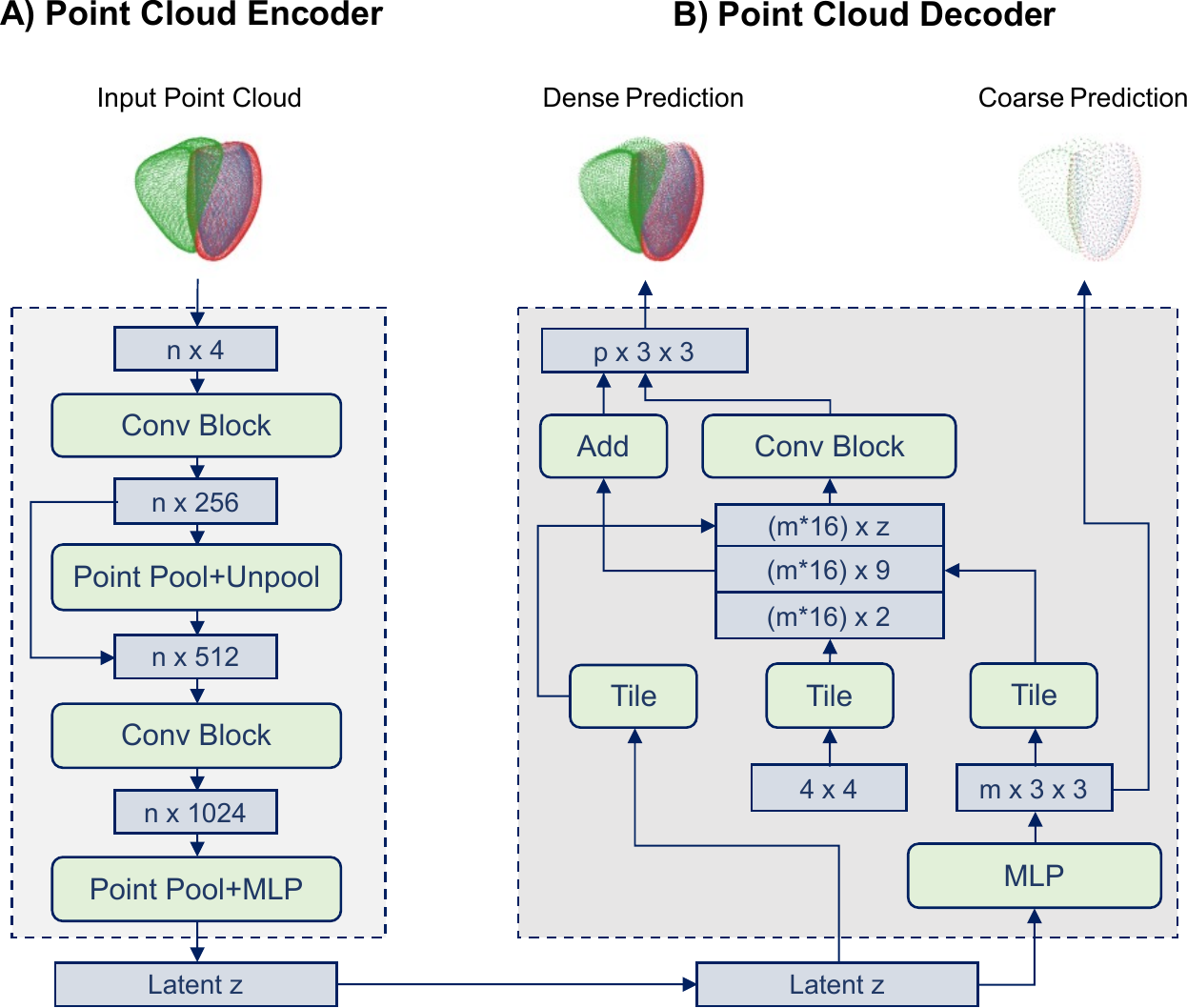}}
	\caption{Architecture of the proposed PCD-Net. Multi-class point clouds consisting of $n$ points are used as network inputs. The PCD-Net aims to predict both a low-resolution point cloud with $m$ points (as an intermediate representation of global shape to facilitate network training) and a high-resolution point cloud with $p$ points (as a final dense output prediction), where $m \ll p$. The architecture consists of an encoder (A) and a decoder (B) connected by a low-dimensional latent space with point cloud-based deep learning components employed throughout the network.}
	\label{fig:network_architecture}
\end{figure}

The PCD-Net architecture follows an encoder-decoder structure (Fig.~\ref{fig:network_architecture}) inspired by previous work on point cloud-based deep learning with non-medical data \cite{qi2017pointnet,qi2017pointnet++,yang2017foldingnet,yuan2018pcn}.

It receives $n\times 4$-dimensional multi-class point cloud representations of the biventricular anatomies as inputs, where $n$ refers to the number of points in the point cloud and $4$ to the x,y,z coordinates and class label (LV endocardium, LV epicardium, or RV endocardium) of each point. The inputs are then passed through the encoder, which consists of two stacked PointNets \cite{qi2017pointnet} followed by a multi-layer perceptron (MLP). It outputs a latent space vector of length $z$, which is then fed into the decoder \cite{yuan2018pcn}. Here, a MLP is used to first create a coarse output point cloud with the aim of facilitating the training process by predicting the heart shape on a global level with low resolution. It is represented as a $m \times 3 \times 3$ tensor where $m$ refers to the number of points and the two $3$s correspond to the number of classes and the x,y,z coordinates of each point, respectively. The coarse output point cloud and the latent space vector are then combined with a set of $4 \times 4$ point patches in an operation inspired by FoldingNet \cite{yang2017foldingnet} to create the final dense output point cloud. Similar to the coarse output, it is represented as a $p \times 3 \times 3$ vector but with $p \gg m$, \emph{i.e.} a considerably higher resolution to adequately incorporate both local and global shape information.

\subsection{Training and Implementation}
We choose a loss function for network training that consists of two separate loss terms, $L_{coarse}$ and $L_{dense}$, and a weighting parameter $\alpha$ \cite{yuan2018pcn}:

\begin{equation}
L_{total} = \displaystyle\sum_{i = 1}^{C} \big(L_{coarse, i} + \alpha * L_{dense, i}\big).
\label{eq_loss}
\end{equation}

Here, $C$ refers to the number of cardiac substructures in the anatomy point clouds and is set to 3 in this work. $L_{coarse}$ and $L_{dense}$ evaluate the prediction error of the coarse and dense output points clouds with respect to the same dense ground truth point cloud. We use low $\alpha$ values of 0.01 at the beginning of training to put more emphasis on learning an accurate low-resolution prediction, before steadily increasing it to larger values of 5.0 as training progresses to put more focus on local prediction accuracy in addition to the global one. For both loss terms, we select the symmetric Chamfer distance between the respective predicted point cloud and ground truth point cloud as an approximate surface-to-surface distance metric on point cloud data. We train all networks using the Adam optimizer with a batch size of 8 on a GeForce RTX 2070 Graphics Card with 8~GB memory, until no improvement on the validation dataset is found for 10000 steps. We split the data into 70\% train data, 5\% validation data, and 25\% test data.

\section{Experiments}

\subsection{Deformation Prediction}
\label{ssec:exp_deformation_pred}

\begin{figure}[t]
	\begin{minipage}[b]{1.0\linewidth}
		\centering
		\centerline{\includegraphics[width=1.0\textwidth]{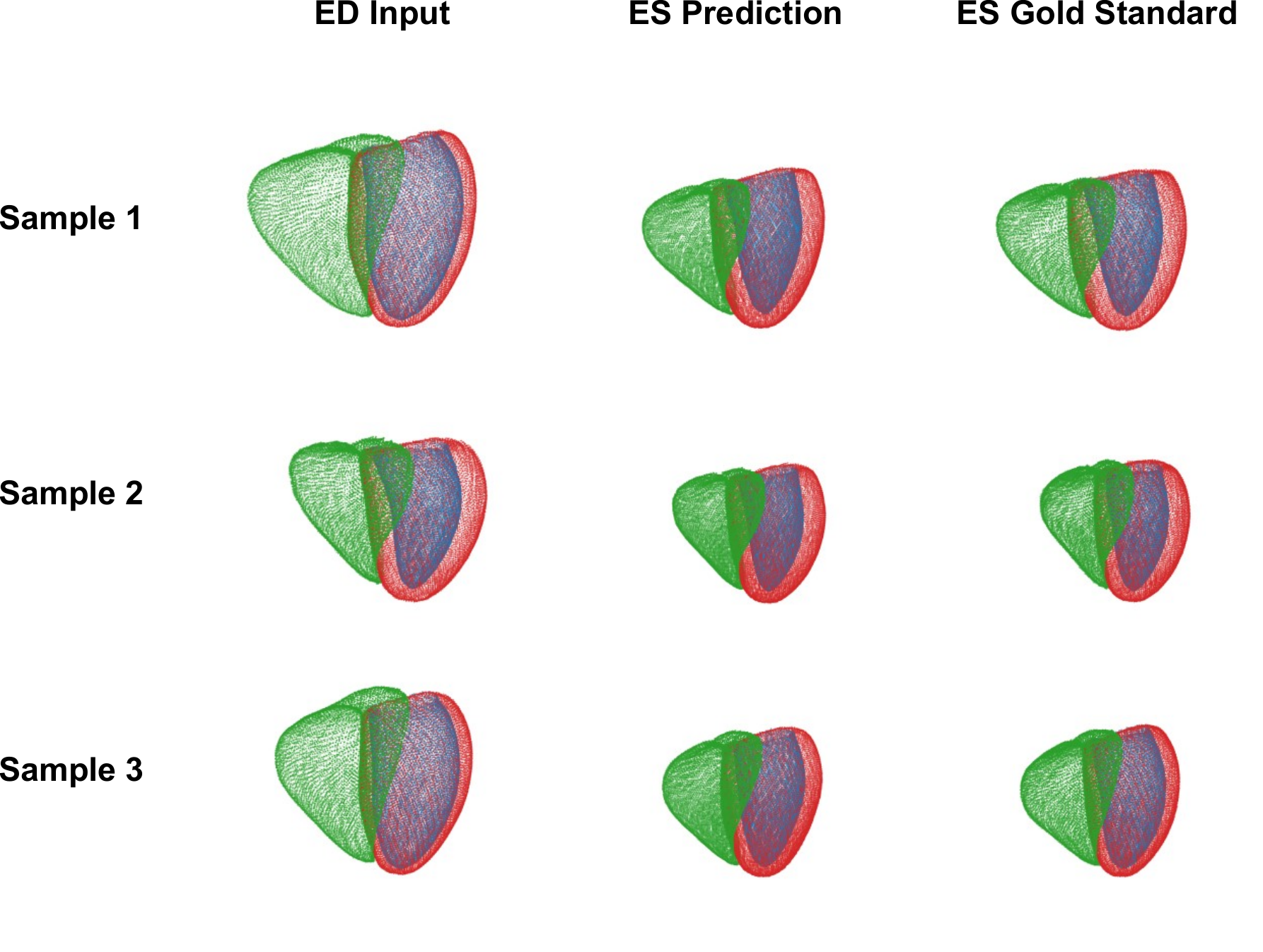}}
	\end{minipage}
	\caption{Qualitative prediction results of three sample cases for cardiac contraction (ED to ES).}
	\label{fig:qual_results_ED_to_ES}
\end{figure}

\begin{figure}[htbp]
	\begin{minipage}[b]{1.0\linewidth}
		\centering
		\centerline{\includegraphics[width=1.0\textwidth]{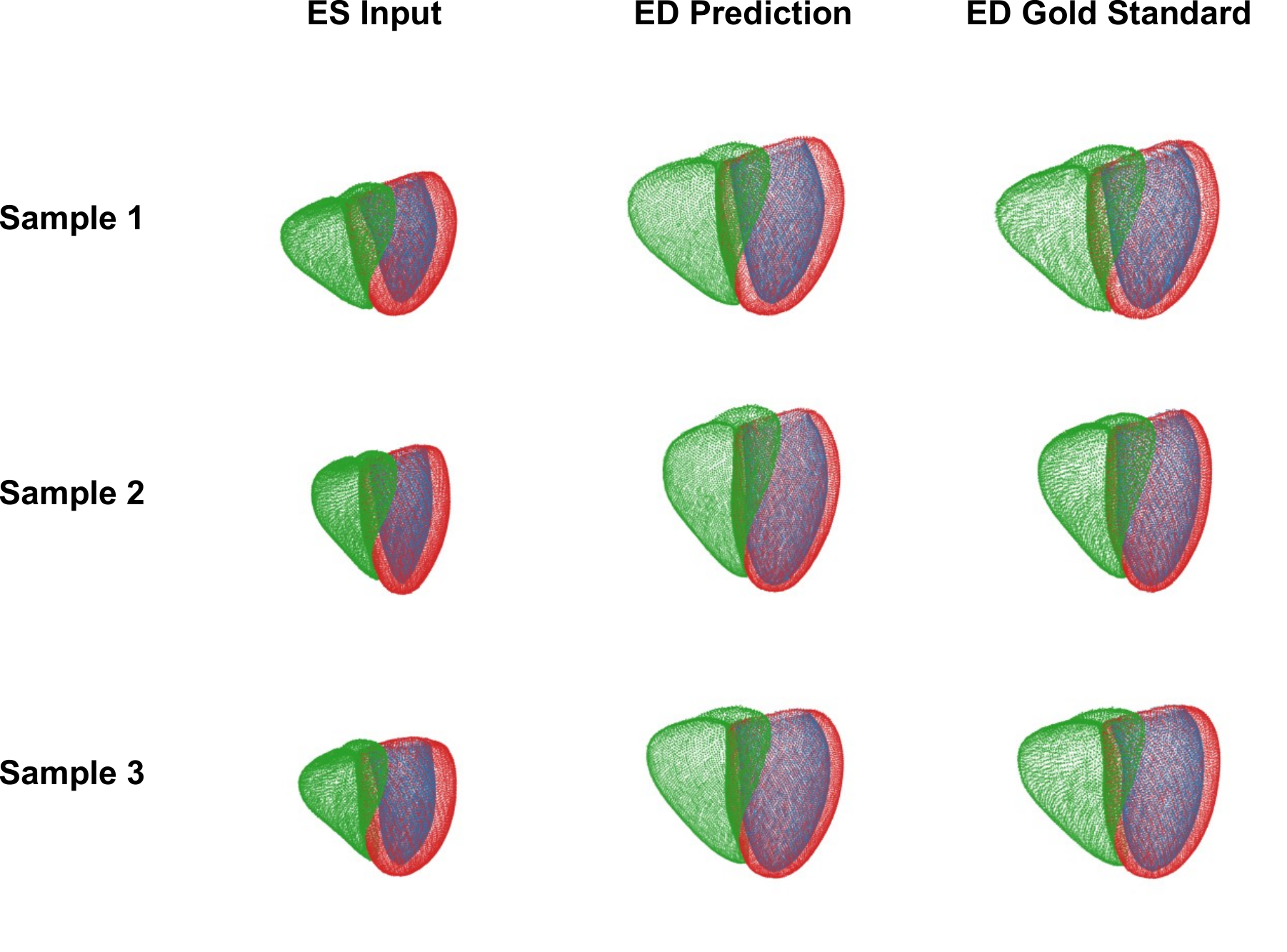}}
	\end{minipage}
	\caption{Qualitative prediction results of three sample cases for cardiac relaxation (ES to ED).}
	\label{fig:qual_results_ES_to_ED}
\end{figure}

In our first experiment, we assess the general ability of the PCD-Net to accurately model 3D cardiac contraction and relaxation. To this end, we train two separate PCD-Nets, one to predict ES shapes from ED shapes as a representation of contraction, and one to predict ED shapes from ES shapes to model relaxation. The input point cloud, the point cloud predicted by the network, and the gold standard point cloud of three sample cases from the test dataset are depicted in Fig.~\ref{fig:qual_results_ED_to_ES} and Fig.~\ref{fig:qual_results_ES_to_ED} for cardiac contraction and relaxation respectively.

We observe a good alignment between predicted and gold standard point clouds on both a local and global level for all cardiac substructures and a variety of different input shapes. Contraction and relaxation are similarly well captured.

To quantify the prediction performance of both PCD-Nets, we calculate the Chamfer distances between the corresponding predicted and gold standard point clouds in the test dataset separately for each cardiac substructure and prediction direction and report the results in Table~\ref{tab:quan_pred_results}.

\begin{table}[!ht]
    \caption{Quantitative prediction results of the PCD-Net for both cardiac contraction and relaxation.}
    \def\arraystretch{1.5}\tabcolsep=3pt
    \centering
    \begin{tabular}{p{30pt}p{35pt}p{75pt}p{80pt}}
        \hline
        Input Phase & Predicted Phase & Class & Chamfer Distance (mm) \\
        \hline
        
         \multirow{3}*{ED}  &  \multirow{3}*{ES}    & LV endocardium  & 1.24 ($\pm$0.56)    \\
           &                            & LV epicardium   & 1.32 ($\pm$0.60)    \\
            &                            & RV endocardium  & 1.48 ($\pm$0.66)    \\
             
        \hline
         \multirow{3}*{ES} & \multirow{3}*{ED}    & LV endocardium  & 1.55 ($\pm$0.68) \\
         &        & LV epicardium   & 1.53 ($\pm$0.80)   \\
         &        & RV endocardium  & 1.83 ($\pm$0.82)   \\
             
        \hline
        
        \multicolumn{4}{@{}l}{Values represent mean ($\pm$ standard deviation (SD)) in all cases.}
    \end{tabular}
    \label{tab:quan_pred_results}
\end{table}

We find average Chamfer distances below or comparable to the voxel resolution of the underlying short-axis images ($1.8 \times 1.8 \times 8.0~\mbox{mm}^3$) for all cardiac substructures and both prediction directions. Errors are slightly higher for cardiac relaxation than for contraction.

\subsection{Clinical Evaluation}

We next analyze the PCD-Net's predictive ability from a clinical perspective. We first focus on the cardiac anatomy at ED and ES separately and select three widely used clinical metrics, the LV volume, RV volume, and LV mass, to assess the quality of the biventricular shapes predicted by our network. We calculate the three metrics for both the predicted and gold standard anatomies of the test dataset and report their respective population-wide mean and standard deviation in Table~\ref{tab:clinical_anatomy_metrics}. In addition, we compute the percentage differences between the metric values of the corresponding predicted and gold standard anatomies of each subject in order to obtain a case-specific measure of prediction performance (Table~\ref{tab:clinical_anatomy_metrics}).
  
\begin{table}[!h]
    \caption{Clinical cardiac anatomy metrics for both predicted and gold standard point clouds.}
    \def\arraystretch{1.5}\tabcolsep=3pt
    \centering
    \begin{tabular}{p{17pt}p{21pt}p{55pt}p{35pt}p{35pt}p{50pt}}
        \hline
        Input Phase & Output Phase & Clinical Metric & Gold Standard & Prediction & Per-case Difference (\%) \\
        \hline
        \multirow{3}*{ED} & \multirow{3}*{ES}    & LV ES vol (ml)  & 59 ($\pm$12)  & 58 ($\pm$14) & 9.25 ($\pm$6.92) \\
           &                        & RV ES vol (ml)  & 75 ($\pm$17)  & 75 ($\pm$19) & 8.80 ($\pm$7.23) \\
            &                        & LV mass (g)  & 105 ($\pm$28)  & 104 ($\pm$27) & 7.85 ($\pm$7.10) \\

        \hline
        \multirow{3}*{ES} &\multirow{3}*{ED}    & LV ED vol (ml)  & 134 ($\pm$26)  & 140 ($\pm$29) & 9.90 ($\pm$7.06) \\
            &                     & RV ED vol (ml)  & 158 ($\pm$31)     & 165 ($\pm$34) & 8.06 ($\pm$5.78) \\
            &                      & LV mass (g)  & 105 ($\pm$28)  &  112 ($\pm$32) & 10.48 ($\pm$8.70) \\
        \hline
        \multicolumn{6}{@{}l}{Values represent mean ($\pm$ SD) in all cases.}
    \end{tabular}
    \label{tab:clinical_anatomy_metrics}
\end{table}

We find similar mean and standard deviation values between the predicted and gold standard populations for all clinical metrics, cardiac substructures, and prediction directions. The average per-case percentage differences between the respective predicted and gold standard scores are approximately 9\% across all metrics.

Apart from purely anatomy-based metrics limited to individual phases of the cardiac cycle, we next investigate the PCD-Net's prediction performance in terms of widely used cardiac function metrics. To this end, we compute the LV and RV ejection fraction (EF) for both the gold standard and predicted anatomies of the test dataset and present the results in Table~\ref{tab:clinical_function_metrics}.

\begin{table}[!h]
    \caption{Clinical cardiac function metrics for both predicted and gold standard point clouds.}
    \def\arraystretch{1.5}\tabcolsep=3pt
    \centering
    \begin{tabular}{p{17pt}p{21pt}p{50pt}p{35pt}p{35pt}p{50pt}}
        \hline
        Input Phase & Output Phase & Clinical Metric & Gold Standard & Prediction & Per-case Difference (\%) \\
        \hline
        \multirow{2}*{ED} & \multirow{2}*{ES}    & LV EF (\%)  & 58 ($\pm$7)  & 57 ($\pm$7) & 6.62 ($\pm$5.26) \\
            &                        & RV EF (\%)  & 55 ($\pm$6)  & 55 ($\pm$5) & 7.24 ($\pm$6.27) \\
        \hline
        \multirow{2}*{ES} &\multirow{2}*{ED}      & LV EF (\%)   & 58 ($\pm$7)     & 57 ($\pm$6)  & 7.67 ($\pm$6.30) \\
            &                      & RV EF (\%)  & 55 ($\pm$6)  & 53 ($\pm$5) & 7.09 ($\pm$5.83) \\
        \hline
        \multicolumn{6}{@{}l}{Values represent mean ($\pm$ SD) in all cases.}
    \end{tabular}
    \label{tab:clinical_function_metrics}
\end{table}

Similar to the results for cardiac anatomy metrics, we observe only small differences in mean and standard deviation scores between the predicted and ground truth populations. Average per-case percentage differences are $\sim$7\% across all metrics.

\subsection{Infarction-specific Deformation Patterns}
\label{ssec:exp_subpopulation_diffs}
\begin{figure}[htbp]
	\begin{minipage}[b]{1.0\linewidth}
		\centering
		\centerline{\includegraphics[width=1.0\textwidth]{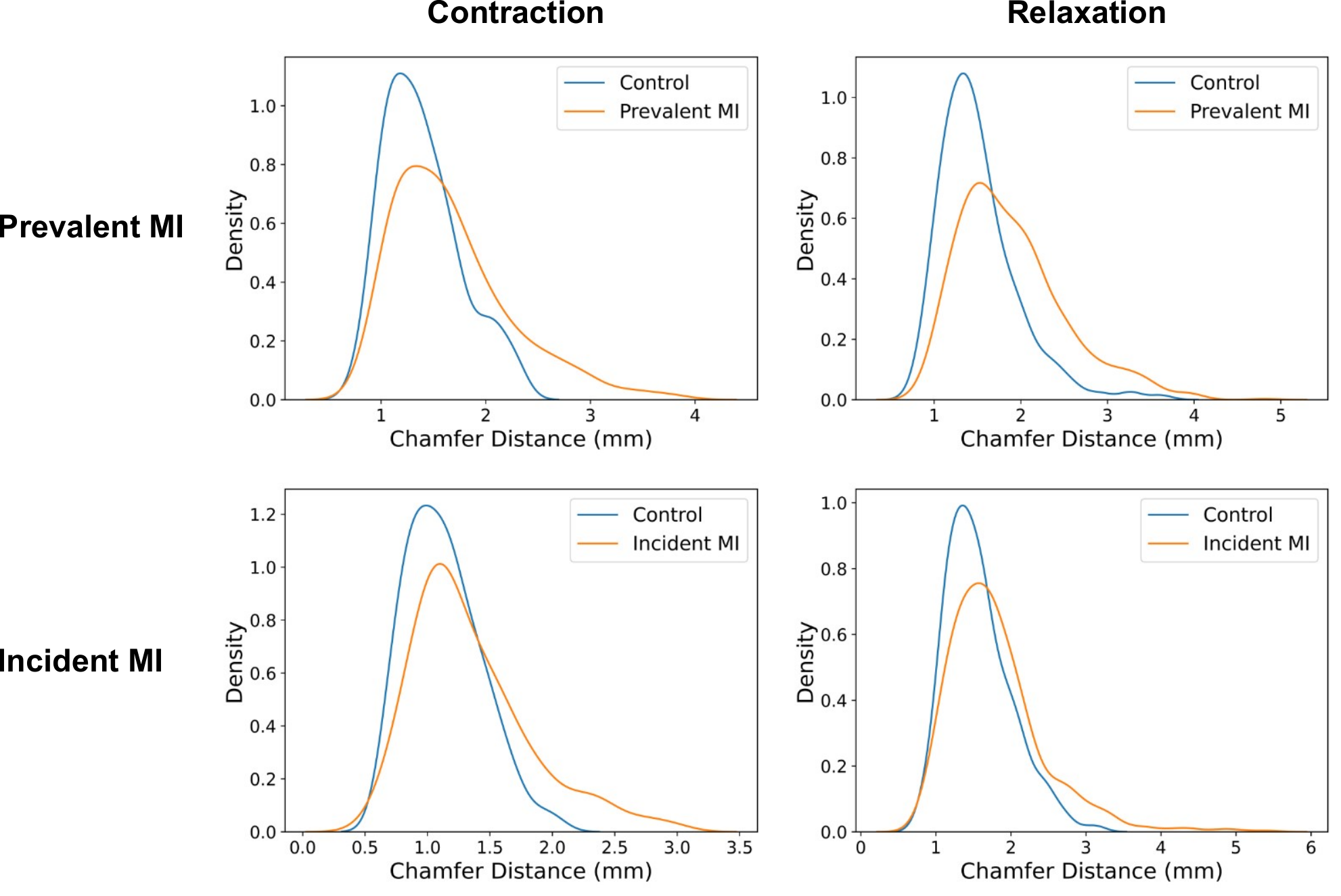}}
	\end{minipage}
	\caption{Distributions of prediction performance of PCD-Nets trained on normal cases and evaluated on both unseen normal cases and MI cases. Results are shown for prevalent MI vs normal cases (upper row) and for incident MI vs normal cases (lower row) for cardiac contraction (left column) and relaxation (right column). p-values of the KS-test for all comparisons are $<$0.0001.}
	\label{fig:subpop_diffs_mi}
\end{figure}

Following the validation on the whole population, we next investigate whether the PCD-Net can learn 3D deformation patterns specific to certain subpopulations. To this end, we first select all cases with a prevalent MI event and combine them with the same number of randomly sampled normal cases to create a test dataset with two subpopulations. We train two PCD-Nets, one to predict cardiac contraction and one to predict cardiac relaxation, on the remaining dataset of control subjects in order to learn a representation of normal 3D cardiac deformation. Then, we apply each of these two pre-trained networks to the unseen test dataset. For each prediction direction, we calculate the Chamfer distances between predicted and gold standard anatomies separately for the normal subpopulation and the pathological subpopulation of the test dataset. This allows us to evaluate whether the network trained exclusively on normal cases performs differently on unseen cases from the same normal subpopulation versus (vs) the diseased cases of the other new subpopulaton. We repeat the same process for incident MI cases and visualize the distributions of Chamfer distances for both prediction directions and MI types in Fig.~\ref{fig:subpop_diffs_mi}.

We find statistically significant differences using the Kolmogorov–Smirnov (KS) test between the prediction performance for MI and normal cases for both MI types and prediction directions. Subpopulation-specific differences are typically larger for prevalent MI vs normal cases than for incident MI vs normal.

\subsection{Prevalent Myocardial Infarction Detection}
\label{ssec:exp_mi_detection}
Given the differences in prediction performance of the PCD-Net between normal and MI cases (Sec.~\ref{ssec:exp_subpopulation_diffs}), we next analyze whether these learned subpopulation-specific 3D deformation patterns can be directly used to detect MI cases. To this end, we first extract the latent space representations of all normal and prevalent MI cases in our test dataset by passing the respective input point clouds through the encoder of the pre-trained PCD-Nets. The extracted values are low-dimensional representations of the subject-specific complex 3D deformation patterns. We then train a logistic regression classifier for the binary detection of prevalent MI vs normal cases using the individual latent space components as independent variables. We execute this procedure separately with the networks trained for cardiac contraction and relaxation, and report the averaged results of 10-fold cross validation experiments in Table~\ref{tab:prevalent_mi_detection} in terms of commonly used binary classification metrics. As a clinical benchmark, we first select EF of the LV, RV, and both ventricles, as these metrics also evaluate cardiac function based on ED and ES and are the primary image-based biomarkers to detect MI in clinical practice. We then train separate logistic regression models with the respective EF values as the independent variable on the same dataset using 10-fold cross validation experiments (Table~\ref{tab:prevalent_mi_detection}).

\begin{table}[!h]
    \caption{Prevalent MI prediction results of regression models with different independent variables.}
    \def\arraystretch{1.5}\tabcolsep=3pt
    \centering
    \begin{tabular}{p{70pt}p{30pt}p{30pt}p{30pt}p{30pt}p{25pt}}
        \hline
         Input & Accuracy & AUROC & F1-Score & Precision & Recall \\
        \hline
         LV EF         & 0.6005   & 0.6624  & 0.5683  & 0.6207 & 0.5373  \\
         RV EF         & 0.5828  & 0.6181  & 0.5559  & 0.6009  & 0.5338  \\
         LV+RV EF       & 0.5964   & 0.6524  & 0.5661   & 0.6128  & 0.5374  \\
         \hline
         Latent z (ES to ED)  & 0.6830  & 0.7387  & 0.6747   & \textbf{0.6934}  & 0.6665  \\
         Latent z (ED to ES)  & \textbf{0.6835}   & \textbf{0.7498}  & \textbf{0.6795}   & 0.6877 & \textbf{0.6817}  \\
        \hline
        \multicolumn{5}{@{}l}{Values represent the mean in all cases.}
    \end{tabular}
    \label{tab:prevalent_mi_detection}
\end{table}

We find that the latent space representations learned by the PCD-Nets outperform all clinical benchmarks for both prediction directions and across all classification metrics. Compared to the best clinical benchmark LV EF, the AUROC scores improve by 13\% for contraction and 12\% for relaxation.

\subsection{Incident Myocardial Infarction Prediction}
\label{ssec:exp_incident_mi_pred}
In addition to detecting prevalent MI events, we also investigate the utility of the PCD-Net's latent space representations for the prediction of incident MI outcomes. We therefore repeat the same procedure of the prevalent MI detection for the binary prediction of incident MI (Table~\ref{tab:incident_mi_prediction}).

\begin{table}[!h]
    \caption{Incident MI prediction results of regression models with different independent variables.}
    \def\arraystretch{1.5}\tabcolsep=3pt
    \centering
    \begin{tabular}{p{70pt}p{30pt}p{30pt}p{30pt}p{30pt}p{25pt}}
        \hline
         Input & Accuracy & AUROC & F1-Score & Precision & Recall \\
        \hline
         LV EF         & 0.5806   & 0.6347  & 0.5437  & 0.6008 & 0.5136  \\
         RV EF         & 0.5145 & 0.5541  & 0.4740  & 0.5336 & 0.4578  \\
         LV+RV EF       & 0.5957  & 0.6248  & 0.5724   & 0.6112  & 0.5542  \\
         \hline
         Latent z (ES to ED)  & 0.6138   & 0.6635  & 0.6045   &  0.6197  & 0.6004 \\
         Latent z (ED to ES)  & \textbf{0.6349}   & \textbf{0.6765}  & \textbf{0.6306}   & \textbf{0.6373} & \textbf{0.6362}  \\
        \hline
        \multicolumn{5}{@{}l}{Values represent the mean in all cases.}
    \end{tabular}
    \label{tab:incident_mi_prediction}
\end{table}

Similar to the results for prevalent MI, we find a better performance of the latent space representations compared to the clinical benchmarks for both prediction directions and classification metrics. Contraction models are more predictive than relaxation models with 7\% and 5\% improvements in AUROC, respectively, over the best clinical benchmark.

\subsection{Myocardial Infarction Survival Analysis}
\label{ssec:exp_mi_survival_pred}
After a binary disease classification, we next study MI prediction in a more time-resolved manner and perform a survival analysis for incident MI events. We therefore first calculate the time in months between the imaging date and the MI event or censoring date for each case. Hereby, we define 1 January 2021 as our study endpoint and consider all cases with no prior MI as censored on this date. We then use the obtained time-to-event data to run multiple Cox proportional hazard regression models \cite{cox1972regression} with the same respective independent variable inputs of LV EF, RV EF, LV EF and RV EF, latent space of ED to ES, and latent space of ES to ED, as in Sec.~\ref{ssec:exp_mi_detection}. We select the widely used Harrell’s concordance index \cite{harrell1982evaluating} as a normalized score between 0 and 1 to compare the predictive performance.
We evaluate each model by running 10-fold cross validation experiments and report the averaged results in Table~\ref{tab:mi_survival_pred}.

\begin{table}[!h]
    \caption{Incident MI survival prediction results of Cox regression models with different independent variables.}
    \def\arraystretch{1.5}\tabcolsep=3pt
    \centering
    \begin{tabular}{p{90pt}p{100pt}}
        \hline
         Input & Harrell’s concordance index  \\
        \hline
         LV EF         & 0.5961    \\
         RV EF         & 0.5577  \\
         LV+RV EF       & 0.5916     \\
         \hline
         Latent z (ES to ED)  & \textbf{0.6347}    \\
         Latent z (ED to ES)  & 0.6272    \\
        \hline
    \end{tabular}
    \label{tab:mi_survival_pred}
\end{table}

In line with the binary classification results, the latent space inputs outperform all clinical benchmarks. Cardiac relaxation achieved the highest C-index score with a 7\% improvement over the best baseline approach.

\section{Discussion and Conclusion}

In this work, we have presented the PCD-Net as a novel geometric deep learning approach to model 3D cardiac deformation between the extreme ends of the cardiac cycle. 

In our experiments, we have seen that the PCD-Net is capable of accurately predicting biventricular shapes with errors below or near the pixel resolution of the underlying image acquisition. This shows that the network architecture can process high-resolution cardiac anatomy models and extract relevant features for 3D deformation modeling. It also demonstrates the PCD-Net's ability to directly work with multi-class point clouds in a single modeling setup, taking deformation information from multiple anatomical substructures and their interactions into account.

Our prediction results also indicate that the biventricular shape information at one end of the cardiac cycle should be sufficient to predict the shape at the other end with reasonable accuracy in cases with normal cardiac function. This is further corroborated by the small population-wide and per-case differences in clinical metrics between gold standard and predicted anatomies. It is achieved despite the fact that no cardiac shapes between ED and ES, subject characteristics such as sex or age, or any other deformation-specific parameters, such as biophysical properties, are supplied to the network. While this puts an inherent limitation on the prediction accuracy attainable with the current setup, we believe that the PCD-Net can be expanded further to include such information. Biophysical properties could for example be added as extra per-point features in the input point cloud in a similar way to the current class information, while a recurrent neural network setup in the latent space could allow the incorporation of other cardiac phases. By successfully predicting both cardiac contraction and relaxation, we have also shown that the proposed approach is flexible enough to capture different motion patterns with the same architectural setup. In addition, the small Chamfer distances achieved by our network indicate that it is not only capable of processing shape information on a global level but also on a local one. This is an important property since localized shape biomarkers have been associated with improved prognostic quality over global ones for a variety of cardiovascular diseases \cite{acero2022understanding,suinesiaputra2017statistical}. All prediction results were accomplished on a large dataset of over 10,000 subjects, which demonstrates the PCD-Net's high robustness and scalability.

We have also found that the PCD-Net can learn 3D deformation patterns specific to certain subpopulations. While such differences are routinely taken into account in clinical practice when assessing cardiac function with single-valued biomarkers and have also been observed in some previous 2D models of cardiac mechanics \cite{Ossenberg-Engels2019}, the PCD-Net can capture considerably more complex and subtle differences by analyzing full 3D shapes. At the same time, it extracts the key aspects of this deformation through a small number of components in an interpretable low-dimensional latent space.
This allows the PCD-Net's latent space to outperform current single-valued clinical benchmarks in multiple pathology classification and survival prediction tasks by considerable margins which is crucial for clinical acceptance and demonstrates the approach's downstream versatility. The results are achieved  while relying on the same basic regression model and despite the latent space representation only capturing general-purpose 3D deformation patterns of normal cases without optimizing for specific downstream tasks during training. We believe that such a combined task-specific training setup could be easily integrated into the given network structure as a separate classification branch and would likely lead to further performance increases and a more targeted latent space. The small performance differences between contraction and relaxation indicate that both contain similarly important pathology-specific patterns. In addition, the latent space-based regression analysis offers a high degree of interpretability, as the effect of individual regressors on the outcome variable can be traced back to 3D shape differences via the network's encoder and decoder. While we did not include any subject characteristics for the outcome prediction task in this work, the presented setup allows for a straightforward addition of such information. Since the PCD-Net is trained to capture normal 3D deformation on a large population, its predictions can also be used to obtain a personalized output shape expected under normal conditions for each case, which can then be compared to the acquired shape. Any deviations between the two shapes might then offer an explainable and easily visualizable way to detect abnormalities. Such normal deformation patterns can also be analyzed for certain subpopulations to further improve performance and understanding. 

While the obtained results corroborate the increased utility of modeling full 3D deformation on three exemplary tasks, we believe that the approach can be readily applied to other cardiovascular risk factors or pathologies associated with an analysis of cardiac deformations.

\section*{Acknowledgments}
This research has been conducted using the UK Biobank Resource under Application Number ‘40161’. The work of M. Beetz is supported by the Stiftung der Deutschen Wirtschaft (Foundation of German Business). A. Banerjee is a Royal Society University Research Fellow and is supported by the Royal Society Grant No. URF{\textbackslash}R1{\textbackslash}221314. The work of A. Banerjee and V. Grau is supported by the British Heart Foundation (BHF) Project under Grant PG/20/21/35082. The work of V. Grau is supported by the CompBioMed 2 Centre of Excellence in Computational Biomedicine (European Commission Horizon 2020 research and innovation programme, grant agreement No. 823712).

\begin{table*}[!htbp]
    \caption{Group of codes and diseases of UK Biobank field ID 20002. Subjects without any of the listed pathologies were selected as control cases in this work.}
    \def\arraystretch{1.5}\tabcolsep=3pt
    \centering
    \begin{tabular}{|p{25pt}p{160pt}|p{25pt}p{160pt}|}
        \hline
        Code & Meaning & Code & Meaning\\
        \hline
        1065 & Hypertension & 1286 & Depression \\
        1066 & Heart/cardiac problem & 1412 & Bronchitis \\
        1067 & Peripheral vascular disease & 1471 & Atrial fibrillation \\
        1072 & Essential hypertension & 1472 & Emphysema \\
        1073 & Gestational hypertension/pre-eclampsia & 1473 & High cholesterol \\
        1074 & Angina & 1483 & Atrial flutter \\
        1075 & Heart attack/myocardial infarction & 1484 & Wolff Parkinson white/WPW syndrome \\
        1076 & Heart failure/pulmonary odema & 1485 & Irregular heart beat \\
        1077 & Heart arrhythmia & 1486 & Sick sinus syndrome \\
        1078 & Heart valve problem/heart murmur & 1487 & SVT/supraventricular tachycardia \\
        1079 & Cardiomyopathy & 1491 & Brain haemorrhage \\
        1080 & Pericardial problem & 1492 & Aortic aneurysm \\
        1081 & Stroke & 1496 & Alpha-1 antitrypsin deficiency \\
        1086 & Subarachnoid haemorrhage & 1531 & Post-natal depression \\
        1087 & Leg claudication/intermittent claudication & 1583 & Ischaemic stroke \\
        1088 & Arterial embolism & 1584 & Mitral valve disease \\
        1111 & Asthma & 1585 & Mitral regurgitation/incompetence \\
        1112 & Chronic obstructive airways disease/COPD & 1586 & Aortic valve disease \\
        1113 & Emphysema/chronic bronchitis & 1587 & Aortic regurgitation/incompetence \\
        1220 & Diabetes & 1588 & Hypertrophic cardiomyopathy \\
        1221 & Gestational diabetes & 1589 & Pericarditis \\
        1222 & Type 1 diabetes & 1590 & Pericardial effusion \\
        1223 & Type 2 diabetes & 1591 & Aortic aneurysm rupture \\
        1262 & Parkinson’s disease & 1592 & Aortic dissection \\
        1263 & Dementia/Alzheimer’s/cognitive impairment & & \\
        \hline
        \multicolumn{4}{@{}l}{}
    \end{tabular}
    \label{tab:uk_biobank_cardiac_disease_codes}
\end{table*}

\bibliographystyle{IEEEtran}
\bibliography{refs}

\end{document}